\documentclass
[amssymb,prl,aps,twocolumn,floatfix,tightenlines,superscriptaddress]{revtex4-1}
\usepackage{graphicx}% Include figure files
\usepackage{color}
\usepackage{eurosym}
\usepackage{dcolumn}% Align table columns on decimal point
\usepackage{bm}% bold math
\usepackage{subfigure}
\usepackage{setspace}
\graphicspath{{./images/}}
\usepackage{amsmath}

\begin{document}

\title{Four wave mixing in 3C SiC Ring Resonators}

\author{Francesco Martini}
\affiliation{School of Physics and Astronomy, University of Southampton, Southampton, SO17 1BJ, United Kingdom}
\author{Alberto Politi}
\email{A.Politi@soton.ac.uk}
\affiliation{School of Physics and Astronomy, University of Southampton, Southampton, SO17 1BJ, United Kingdom}

\begin{abstract}
We demonstrate frequency conversion by four wave mixing at telecommunication wavelengths using an integrated platform in 3C SiC. The process was enhanced by high-Q and small modal volume ring resonators, allowing the use of mW-level CW powers to pump the nonlinear optical process. We retrieved the nonlinear refractive index $n_{2}=(5.31\pm 0.04)\times 10^{-19} m^{2}/W$ of 3C SiC and observed a signal attributed to Raman gain in the structure.
\end{abstract}

\maketitle

Third order nonlinear effects are relevant for a broad variety of optical process in integrated structures, ranging from wavelength conversion\cite{Turner:08,Absil:00,Foster:07}, amplification\cite{foster2006broad} and self phase modulation\cite{PhysRevA.17.1448} to generation of non-classical state of light\cite{Azzini2013}. These effects were widely studied in Silicon (Si)\cite{leuthold2010nonlinear}, due to its high $\chi^{(3)}$ susceptibility and scalability, and Silicon Nitride\cite{Levy2009} (Si$_{3}$N$_{4}$), due to its low propagation loss and wide bandgap. Recently, photonic platform in Aluminium Nitride\cite{Jung:13} and Diamond\cite{Hausmann2014} have reached competitive results. However, the development of quantum technologies has risen interest in new materials that are able to integrate different capabilities\cite{OBrien2009}, like the presence of quantum emitters\cite{Calusine2014} and a non-centrosymmetric crystalline structure\cite{Yamada2014}. In addition to meet these requirements, Silicon Carbide (SiC) offers a wide bandgap (2.3 eV), a high refractive index (2.6) and high-quality layers are commercially available. Between all the different polytypes, cubic SiC (3C SiC) can be grown heteroepitaxially on top of Si substrates, reducing the fabrication steps required to fabricate integrated structures. Recently, $\chi^{(3)}$ effects were demonstrated in amorphous SiC toroids at telecommunication wavelengths\cite{Lu:14} and in 4H SiC waveguides in the mid infrared wavelengths\cite{cardenas2015optical}. Even though numerical predictions are available for 3C SiC $\chi^{(3)}$ susceptibility \cite{de2017dispersion}, third order nonlinear effects have not been experimentally demonstrated due to difficulties in fabricating sub-$\mu$m photonic devices. Recently, we demonstrated high confinement optical components in suspended 3C SiC\cite{Martini:17}, suitable for exploring optical nonlinearities.

In this letter we report the demonstration of frequency conversion by four-wave mixing (FWM) in 3C SiC ring resonators, fully integrated in an optical circuit composed of grating couplers, bus waveguides and mode converters. We achieved a conversion efficiency of $-72dB$ at the low pump power of $2.9mW$. The absence of two photon absorption at telecom wavelength does not limit the pump intensity, meaning that high conversion efficiencies are possible. The retrieved nonlinear refractive index $n_{2}=(5.31\pm 0.04)10^{-19} m^{2}/W$ is almost twice the value of Si$_{3}$N$_{4}$ and comparable to many materials used for nonlinear photonics.

Optical nonlinearities can be enhanced decreasing the cross-section of waveguides\cite{Foster2008}, in which the small mode area $A_{eff}$ increases the effective nonlinearity $\gamma$ of structures
\begin{equation}\label{eq1}
     \gamma = \frac{n_{2}\omega_{p}}{A_{eff}c} 	 
\end{equation}
where $\omega_{p}$ is the pump frequency and $c$ is the speed of light in vacuum. Thanks to the tight confinement, efficient frequency conversion was demonstrated using millimeter long high-confinement waveguides together with a broad conversion bandwidth\cite{Foster:07}, provided anomalous group velocity dispersion (GVD) is obtained for the structures. 

\begin{figure*}[htbp]
 \centering
  \centering
  \includegraphics[width=0.55\textwidth]{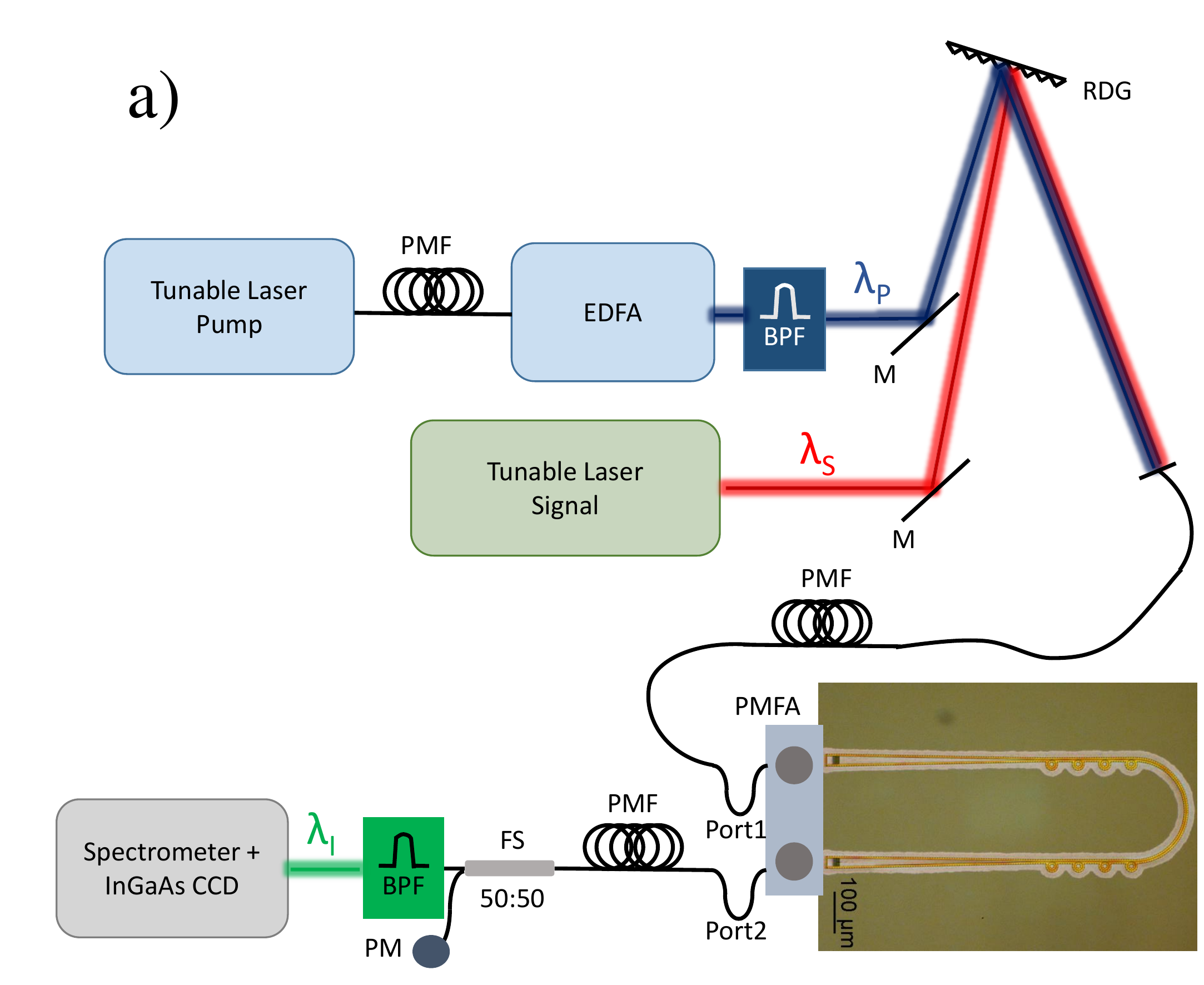}
  \hspace{5.00mm}
  \raisebox{0.8cm}{\includegraphics[width=0.4\textwidth]{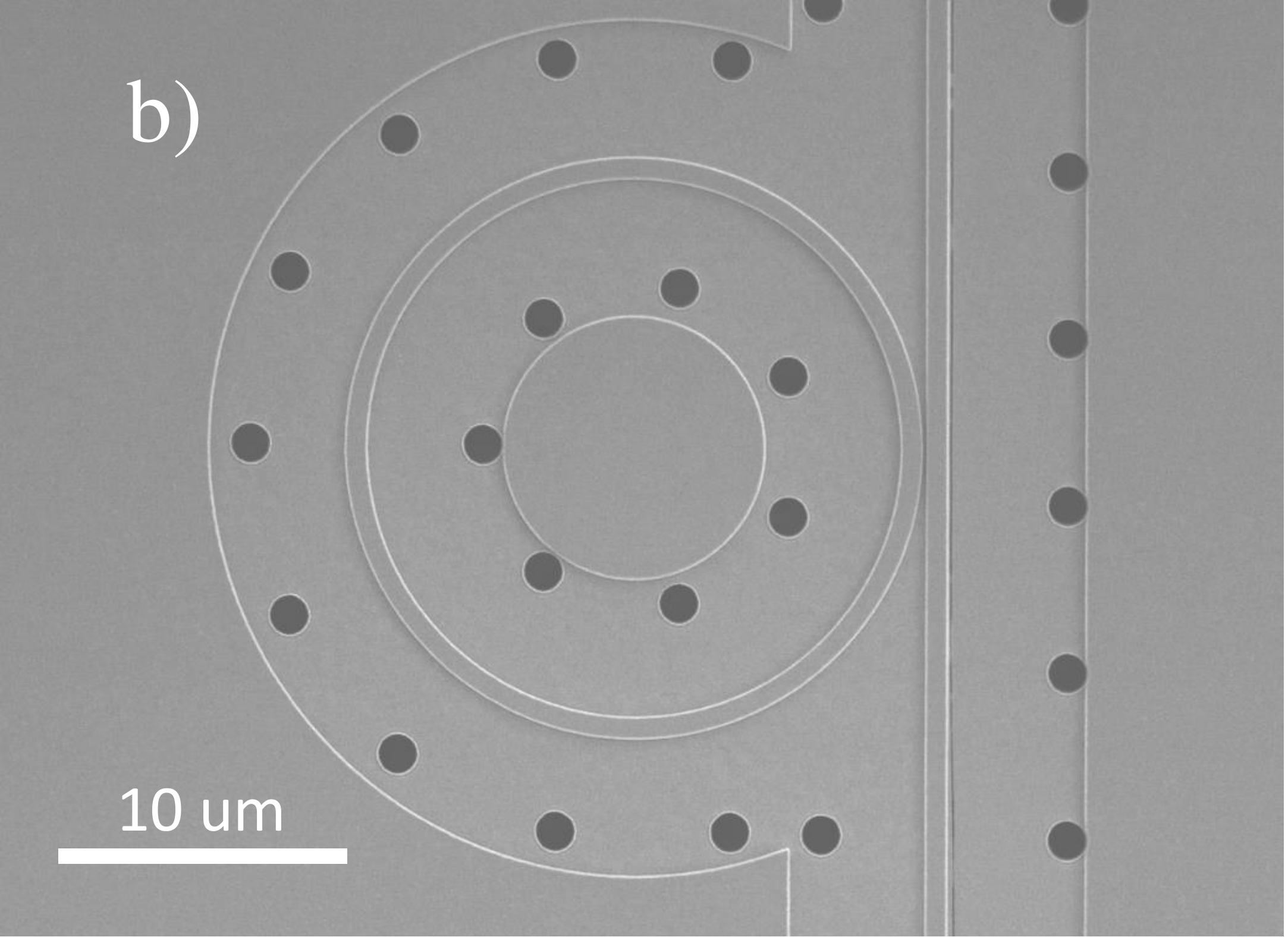}}
  \caption{a) Experimental setup for the characterization of the FWM process, including an optical micrograph of the sample. PMF, Polarization Maintaining Fiber; EDFA, Erbium-Doped Fiber Amplifier; BPF, Band Pass Filter; RDG, Reflective Diffraction Grating; M, Mirror; PMFA, Polarization Maintaining Fiber Array; FS, Fiber Splitter; PM, Power Meter. b) SEM view of the multimode ring resonator.}
  \label{F1}
\end{figure*}

Further improvements in both conversion efficiency and footprint of devices can be achieved using cavity structures, including ring resonators\cite{Absil:00}. When the frequency of light is resonant with the cavity, the intensity of the electromagnetic field is enhanced, increasing the conversion efficiency $\eta$ of the nonlinear process. For a resonator of length $L$ and propagation losses $\alpha$, in which the field is attenuated by a factor $a=\textrm{exp}(-\alpha L/2)$ for a round trip, the field enhancement ($FE$) is dependent by the field coupling from the bus waveguide to the ring $\sigma$, and the transmitted field through the bus waveguide $\tau$ (with $\sigma^2+\tau^2=1$), by
\begin{align}\label{eq2}
	  FE&=\left\lvert  \frac{\sigma}{1-\tau \;\textrm{exp}(-\alpha L + jk L)}  \right\rvert&&
\end{align}
where $k$ is the wavenumber. The maximum value of $FE$ is reached when the field wavelength is resonant with the cavity ($kL=2m\pi$, with $m$ integer) and the resonator is in critical coupling condition ($a=\tau$). Satisfied the previous conditions and in the ideal case of $\sigma<<1$ (low loss regime), the $FE$ can be linked directly to the quality factor (Q) of the cavity 
 \begin{align}\label{eq3}
 	  FE&=\sqrt{\frac{2Q}{k_{p}L}}
 \end{align}
By looking at Eq.\ref{eq1} and \ref{eq3}, an increase in the Q factor and a reduction in the modal volume of the resonator correspond to the enhancement of the field inside the photonic structure, providing increased FMW efficiency.

The SiC device used for FWM experiments was fabricated using two steps of electron beam lithography to define waveguides suspended in air, as reported in our previous work\cite{Martini:17}. The device includes waveguides composed by a $\sim$730$\times 500nm$ structure on top on a $200nm$-thick membrane, resulting in multi-mode operation at telecom wavelength and rings resonators of radius $10\mu m$. In Fig.\ref{F1}.(b) is reported a SEM view of the resonator structure. The ring was close to critical coupling condition with a loaded Q of $\sim$7400 and, due to the propagation loss of $36.6dB/cm$, we used the un-approximated model to describe the FWM process\cite{Absil:00}, in which 
\begin{align}\label{eq4}
     \eta &= \lvert \gamma P_{p} L' \rvert^2 \; FE_{p}^4 \; FE_{s}^2 \; FE_{i}^2 &&\\
	  L'^2&= L^2 \; \textrm{exp}(-\alpha L) \left\lvert \frac{1-\textrm{exp}(-\alpha L + j\Delta k L)}{-\alpha L + j\Delta k L} \right\rvert ^2&&
\end{align}
where $FE_{i}$ is the field enhancement (Eq.\ref{eq2}) for the pump, idler and signal and 
 \begin{align}\label{eq5}
 	  \Delta k=2k_{p}-k_{s}-k_{i}
 \end{align}
is the phase mismatch due to the dispersion of the waveguide. By performing the linear characterization of the sample\cite{rabiei2002polymer} , we retrieved a $FE$ of 4.37 for the pump and signal, and a reduced value of 4.27 for the idler. Even though the wavelengths of the signal, idler and pump met three successive resonances of the cavity, the phase mismatch accumulated in the $\sim 26nm$ bandwidth caused the difference in $FE$ value for the idler.

The experimental setup used for the characterization of the FWM process is depicted in Fig.\ref{F1}.(a) where the pump ($1549.99nm$) and signal ($1563.42nm$) were coupled in the same polarization maintaining fiber using a reflective diffraction grating, additionally providing more than $-45dB$ noise reduction at the idler frequency. The pump power was controlled by a Pritel polarization maintaining erbium-doped fiber amplifier, whose excess noise was attenuated using an additional band pass filter. 
\begin{figure}[htbp]
 \centering
  \includegraphics[clip,trim=2cm 0.8cm 3cm 2cm,width=1\linewidth]{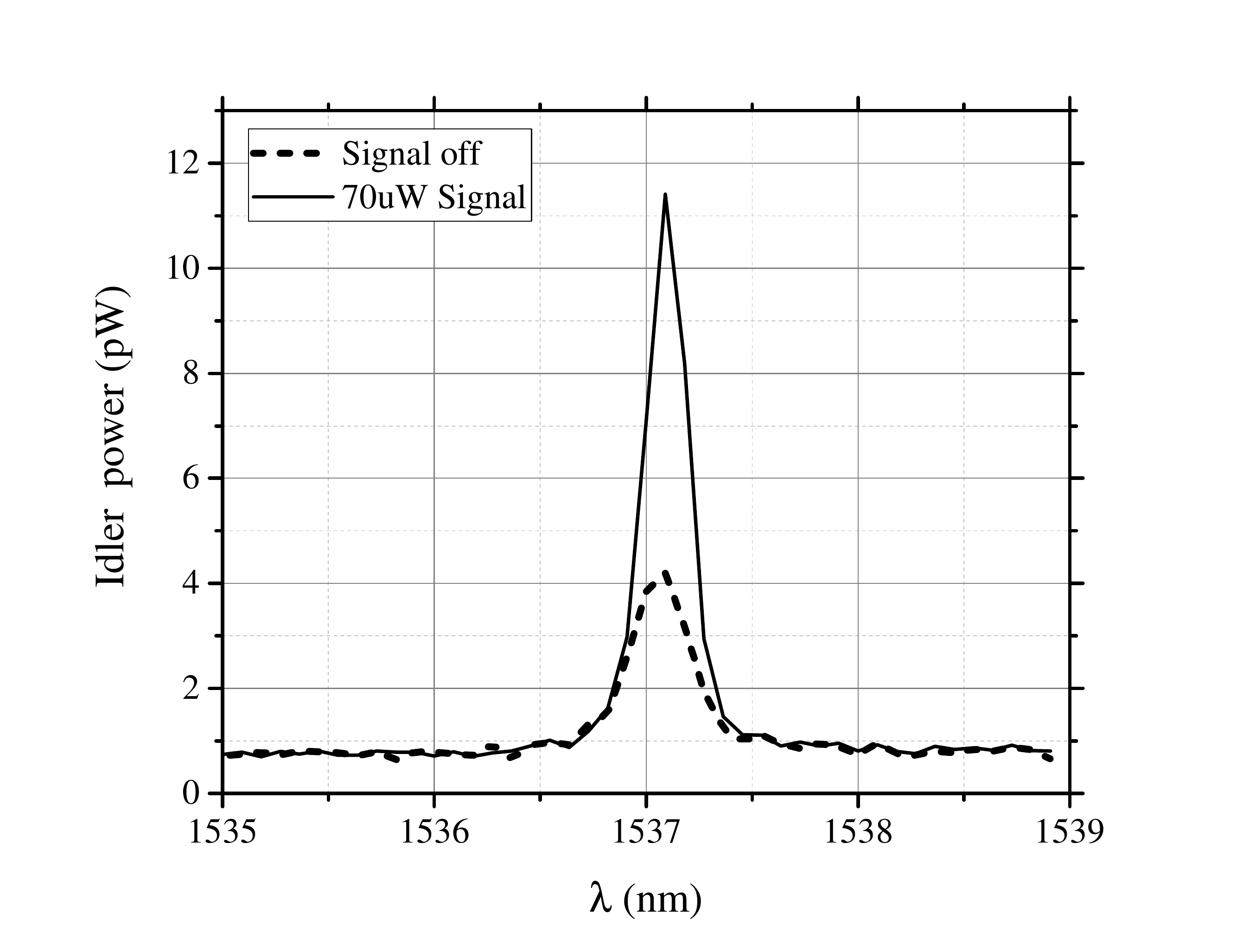}
  \caption{Spectrum of the generated light at the idler wavelength for $1.2mW$ of on-resonance pump power with and without $70\mu W$ of signal input (black and dashed curves, respectively).}
  \label{F2}
\end{figure}
The combined pump and signal waves were injected in the SiC waveguide using a polarization maintaining fiber array and apodized grating couplers (with $\sim -10dB$ coupling efficiency). In the output, the pump and signal were attenuated by the spectrometer and a band pass filter, meanwhile the generated idler intensity was measured with an InGaAs CCD array. A power meter was used to guarantee that the pump and signal were resonant with the ring resonator during the experiment, by minimizing the transmission. 

For a pump power of $1.2mW$, the measured idler power at the ring position is reported in Fig.\ref{F2}. The black curve represents the idler power measured at the ring with both signal ($70\mu W$) and pump on resonance, meanwhile the dashed curve was measured with the signal laser switched off and the pump on resonance. To establish the origin of the light spontaneously generated inside the resonator and extract the non-linear properties of the SiC waveguide, a pump power dependence measurement was performed.
\begin{figure}[htbp]
 \centering
  \includegraphics[width=0.8\linewidth]{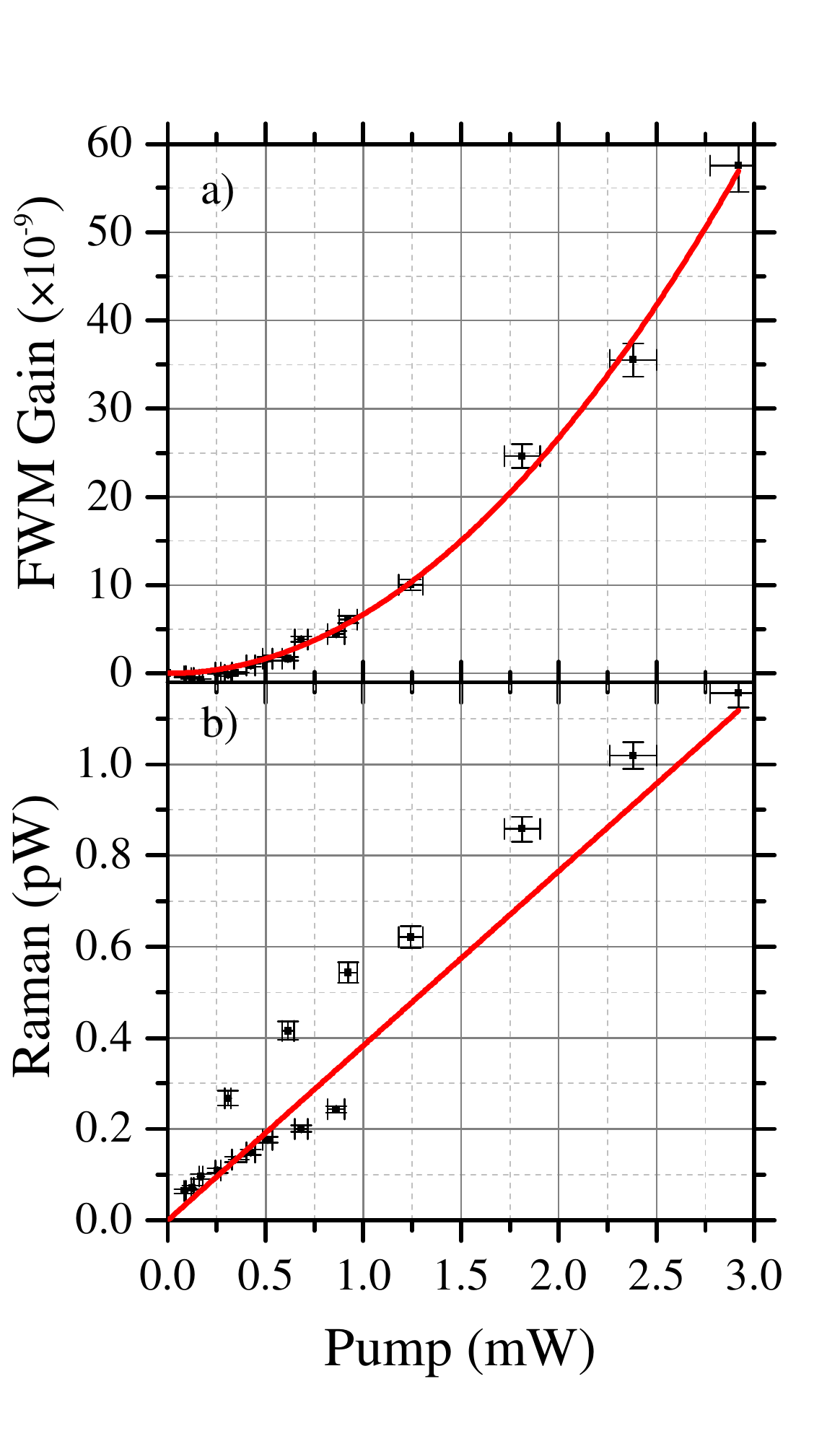}
  \caption{a) Experimental FWM gain (black points) and quadratic fit (red line). b) Raman signal generated in the ring resonator versus resonant pump power (black points) and linear fit (red line).}
  \label{F3}  
\end{figure}
As depicted in the optical micrograph in Fig.\ref{F1}.(a), the sample is composed of a waveguide that couples to eight ring resonators with different radii, each corresponding to a different resonant frequency. This was made to overcome fabrication tolerances and deliver one ring with a resonance within the wavelength tuning range of the pump laser ($\sim 1.5nm$). In order to determine exactly the losses $\beta$ between the fiber array and the ring used for frequency conversion, we performed the experiment twice by inverting the input and output ports of the fiber array. With reference to Fig.\ref{F1}.(a), the pump injected in the sample through the Port 1 (Port2) of the fiber array was attenuated by a factor $\beta_{1}$ ($\beta_{2}$) at the ring position and corresponded to the FWM efficiency $\eta_{1}$ ($\eta_{2}$). Considering $T$ the total transmission between Port 1 and Port 2, and from Eq.\ref{eq4} 
\begin{align*}\label{eq6}
     \eta_{1}/\eta_{2} &= (\beta_{1}/\beta_{2})^2 &&\\
	  T&=\beta_{1}\beta_{2}&&
\end{align*}
we retrieved the input loss values of $-12.61dB$ and $-6.99dB$. This difference in losses between the two experiments are attributed to propagation losses and inhomogeneities deriving from the fabrication process.

The data points of the two experiments are combined in Fig.\ref{F3} by taking in to account the extracted $\beta_i$ values. The FWM gain for different pump powers is reported in Fig.\ref{F3}.(a) where the corresponding values of the generated spontaneous light (Fig.\ref{F3}.b) were subtracted. By fitting Fig.\ref{F3}.(a) with the nonlinear gain of Eq.\ref{eq4}, we retrieved the nonlinearity of the structure $\gamma=3.86\pm 0.03W^{-1}m^{-1}$. The nonlinear susceptibility can be estimated from Eq.\ref{eq1}, in which\cite{Rukhlenko:12} 
\begin{equation}\label{eq7}
     A_{eff} = A_{NL} \iint_{-\infty}^{\infty} S_{z}\;dx dy \bigg/ \iint_{NL} S_{z}\;dx dy \\	 
\end{equation}
with $A_{NL}$ the area of the waveguide cross-section and $S_{z}$ is the Poynting vector parallel to the propagation direction. The absence of a cladding able to participate to the FWM process is accounted in Eq.\ref{eq7}, where only the power flow in the nonlinear ($NL$) SiC cross-section is integrated in the denominator. Using a mode solver, we calculate $A_{eff}=0.558(\mu m)^2$, corresponding to $n_{2}=(5.31\pm 0.04)\times 10^{-19} m^{2}/W$. This value is in partial agreement with the estimated value for bulk crystalline SiC\cite{de2017dispersion} of $4.87\times 10^{-19}m^{2}/W$. 

The spontaneous light generated without signal input is shown in Fig.\ref{F3}.(b) and is fitted using a linear dependence with the pump power. This dependence suggests that this light is generated by Raman scattering, even if in crystalline 3C SiC, Raman signals due to the main vibrational modes\cite{PhysRevB.50.17054} are expected at $\sim 200nm$ away from the pump wavelength. It is possible that the high density of crystalline defects at the SiC-Si interface\cite{anzalone2011advanced} might introduce a complex continuous spectrum that can benefit of the $FE$ in the resonator and generate a Raman field. From the fit to the experimental data we estimated a Raman gain in the ring of $(3.8\pm 0.3)\times 10^{-10} W/ W$. To the best of our knowledge, low energy measurements of heteroepitaxially-grown SiC thin films are not reported in literature and the in-depth analysis of this effect is left for future works.

In conclusion, we demonstrated frequency conversion in a 3C SiC ring resonator. The tight confinement provided a high nonlinearity of the structure of $3.86 W^{-1}m^{-1}$ while the use of the small radius ring resonator enhanced the nonlinear gain of $\sim 50dB$  compared to a straight waveguide of the same length. This is a fundamental step in the development of 3C SiC nonlinear photonic and shows the potential of this platform.

\section*{Funding Information}
This work was supported by the Engineering and Physical Sciences Research Council (EPSRC) EP/P003710/1.

\section*{Acknowledgments}
We acknowledge support from the Southampton Nanofabrication Centre and experimental support by Otto Muskens.

\end{document}